# The MARTE Imaging Spectrometer Experiment: Design and Analysis

Adrian J. Brown[*,1,2], Brad Sutter[2,†] and Stephen Dunagan[2]

[1] *SETI Institute, 515 N. Whisman Rd, Mountain View, CA 94043, USA*

[2] *Ames Research Center, Moffett Field, CA 94035, USA*

## ABSTRACT

We report on the design, operation and data analysis methods employed on the Imaging Spectrometer Instrument as part of the Mars Analog Research and Technology Experiment (MARTE). The Imaging Spectrometer is a hyperspectral scanning pushbroom device sensitive to wavelengths from 400-1000nm. During the MARTE project, the spectrometer was deployed to the Rio Tínto region of Spain. We analyzed subsets of three cores from Rio Tínto using a new band modeling technique. We found most of the MARTE drill cores to contain predominantly goethite, although spatially coherent areas of hematite were identified in core 23. We also distinguished non Fe-bearing minerals that were subsequently analyzed by XRD and found to be primarily muscovite. We present drill core maps including goethite and hematite, and non Fe-bearing minerals.

## KEYWORDS

Mars, Analog, Rio Tinto, Reflectance Spectroscopy, Infrared Spectroscopy

---

[*] corresponding author, email: abrown@seti.org
[†] Now at Jacobs NASA/JSC Mail Code JE23 2224 Bay Area Blvd, Houston, TX 77058

# INTRODUCTION

I n September of 2005, remote subsurface drilling procedures were tested in the Rio Tínto, Spain as a precursor to drilling on Mars. The experiment package was called the Mars Analog Research and Technology Experiment (MARTE) and consisted of a remotely operated drilling platform and associated instruments that examined the retrieved drill core. One of the instruments on board the drilling rig was a Visible-Near Infrared (VNIR: 400-1000nm) Imaging Spectrometer. This paper describes the design and operation of the spectrometer, how the VNIR spectrometer data were analyzed and presents the results of analyzing subsets of three cores from the MARTE project.

# THE MARTE PROJECT

The MARTE project was conceived, planned and executed in 2003-2005 and a remote science experiment took place in September 2005 [*Stoker, et al.*, 2007]. MARTE was conceived as a high definition simulation of remote drilling on the Martian surface in order to identify subsurface life. Several instruments were mounted on a drill rig to remotely analyze the core once it was extracted. In addition to the VNIR imaging spectrometer described herein, the instrument suite included a Raman spectrometer [*Chen, et al.*, 2007], VNIR point spectrometer [*Sutter, et al.*, 2007] and several organic analysis devices [*Stoker, et al.*, 2007].

The MARTE project was conducted in the Río Tinto region of the Iberian pyrite belt in southern Spain as an analog site for Mars. The Río Tinto region is unique due to the unusual preponderance and diversity of iron-rich minerals formed in and on the periphery of the Río Tinto river. The river is colored deep red as a result. This unusual situation is caused by hydrothermal activity linked to the Hercynian orogeny [*Schermerhorn*, 1970]. Hydrothermal waters coursed through Devonian to Carboniferous volcanics and created the subvolcanic Iberian pyrite belt. Later orogenesis has exposed the pyrite to surface water, and microbially mediated oxidation has produced an acid sulfate environment in which minerals such as jarosite, hematite, goethite, barite, and more exotic Fe-oxides, Fe-oxyhydroxides and Fe-sulfate-hydrates have precipitated [*Fernandez-Remolar, et al.*, 2003]. The Río Tinto oxidized Fe mineralogy is similar to that found on Mars, however we do not take this as an indication that Fe oxides on Mars must be microbially mediated. Nevertheless, we do believe the mineral suite available at the Río Tinto serves as an analog in which to test technologies and instrumentation intended for Mars [*Fernández-Remolar, et al.*, 2005].

## MARTE VNIR IMAGING SPECTROMETER

The MARTE imaging spectrometer system shown in Figure 1 was developed at NASA Ames Research Center. The imaging spectrometer was included as part of the drill-core analysis suite on a Mars analog life detection mission – not as a direct indicator of life but in order to give mineralogical context to other instruments.

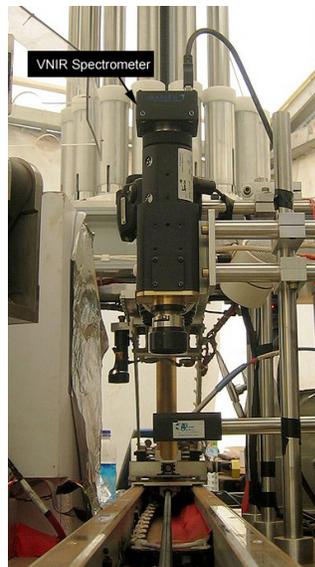

Figure 1. The MARTE imaging spectrometer system mounted on the drill core bench. The vertical black spectrometer housing is approximately 15cm high.

The spectrometer incorporates an imaging spectrograph and monochrome industrial vision CCD camera in a scanning configuration which permits high-resolution imaging of rock core samples with a high degree of spectral information [*Hyvarinen, et al.*, 1998]. The instrument utilizes macroscopic imaging optics and a slit input aperture to sample a line on the object (core). Scanning the core along its axis (normal to the slit) while taking multiple images generates a data "hypercube" in the mode of classic "pushbroom" type remote sensing instruments (e.g. PHILLS, HYDICE [*Kappus, et al.*, 1996; *Davis, et al.*, 2002]).

The spectrograph operates under controlled illumination and a spectral calibration is performed using vapor emission lamps. Radiometric calibration is achieved using a uniform Spectralon calibration target in the object field. A "quick-look" color image is produced at the end of the scan comprised of three wavelength bands that represent normal human perception of blue (480 nm), green (530 nm) and red (660 nm). The instrument operator may specify regions of interest in the quick-look image, and load a request to the hypercube data mining software, which will return reflectance spectra data averaged over these regions.

**Principles of Operation**

The optical system is depicted in Figure 2, with rays tracing the propagation of blue, red, and infrared light (shown as purple in the figure). The objective focuses the target object (in this case the surface of the core) onto a slit input aperture. Light passing through the slit is collimated, refracted by a wedge prism, and directed to a diffraction grating. The first-order diffracted light from the grating is passed through another wedge prism and focused by a second infinite-conjugate-ratio lens to an array detector. A dichroic order-blocking filter (obf) is mounted in front of the CCD to eliminate interference from zero or second order diffraction.

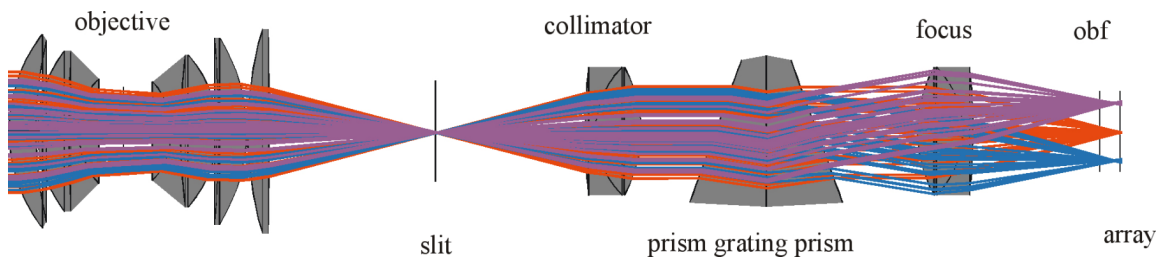

Figure 2. The MARTE imaging spectrometer optical system diagram.

This optical system has the effect of spectrally dispersing the light across one dimension of the array detector. The con-focal collimating lenses inside the spectrograph relay the slit image to the other (spatial "cross scan") dimension of the array detector, providing an image of a line from the object. The second spatial dimension of the hypercube is scanned out by taking sequential spectral frames while linearly translating the spectrograph with respect to the target object at constant focal range in a direction normal to the slit orientation. This was accomplished using the translation mechanism of the MARTE core processing system. The scan parameters (speed, frame rate) were controlled to move the spectrograph a distance equal to the "cross-scan" pixel size in object space in order to have "square" pixels in the hypercube.

|  | Cross scan direction | Along Scan Direction |
| --- | --- | --- |
| Spatial resolution at detector | 8.3 μm (CCD pixel size limited) | 13 μm (slit width limited) |
| Spatial resolution at core (f=17mm, Magnif.=0.25) | 50 μm | 81 μm |
| Spatial sampling interval of scanner |  | 50 μm |

Table 1. MARTE imaging spectrometer system spatial sampling characteristics.

**Design Features**

The imaging characteristics of the spectrograph are determined by several optical design choices. The input slit aperture is 13 micrometers wide. The array detector was oriented with its longer dimension spanning the spectral dimensions of the image plane in order to achieve the maximum spectral range. The array comprises a Sony ICX 415AL/AQ progressive scan CCD chip having 782 by 582 effective pixels, an 8 mm diagonal (Type 1/2'); and unit cell size of 8.3μm by 8.3μm. The spectrograph collimating and focal lenses provide unmagnified imaging of the slit to the array detector in the spatial dimension, therefore the instantaneous spatial pixel dimension is 13 μm along track by 8.3 μm cross track, times the magnification ratio of the objective (Table 1). A fixed-focus (infinite conjugate) objective lens with 17 mm focal length was selected to provide appropriate magnification of the 27 mm core diameter across the full spatial dimension of the array detector (Table 1). This spatial resolution is further elongated in the track direction by the product of scan-speed and camera shutter exposure time, however this effect is negligible for the slow scan speeds used in this experiment.

The spectrograph resolved 580 columns of 50-$\mu$m spatial elements along the input aperture. A 50 $\mu$m spatial size pixel was chosen to be a reasonable fit to expected mineral grain size. The hypercube image length was specified by the length of the scan, configured for rock cores 25 cm long with 1 cm of overlap on either end. The spectrum was sampled by 780 detector rows spanning the 400 to 1000nm wavelength range. The 780 rows were binned into six-pixel averages for a total of 130 final bands. The hypercubes produced in this experiment can be visualized as a stack of monochrome images of increasing wavelength as one "drills down" through the stack. In fact, the vertical data distribution represents the reflectance spectrum for a particular pixel on the core surface. All hypercubes were of identical dimensions - 5221 pixels long by 580 pixels wide and with 130 spectral channels.

The camera well depths and analog-to-digital converters limit signal to 8 bits of discrimination at peak intensity, occurring near the green portion of the spectrum. Reflectance measurements are considerably degraded at the long- and short-wave ends of the spectral range by a combination of: 1) reduced quantum efficiency in silicon and 2) reduced illumination power in the shortwave end of the tungsten-halogen illumination lamp spectrum. To address this problem, an amethyst filter was used to suppress the green peak in the detector response, permitting the use of longer integration times to increase the blue and IR signal strength without saturating the green wavelengths. These effects are depicted in Figure 3. Normalized curves for detector response, lamp illumination (color temp = 3000 K), and filter pass are shown in Figure 3a. Figure 3b shows the spectral response of the instrument resulting from the combination of these effects, illustrating the loss of dynamic range and signal at the ends of the spectrum.

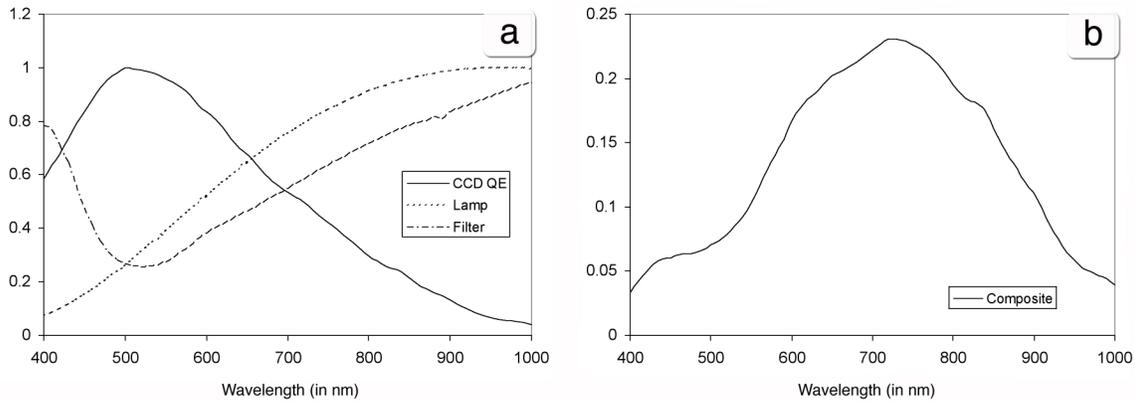

Figure 3. Spectral calibration curves. (a) CCD quantum efficiency, lamp intensity, and filter spectral curve (b) Calculated spectrometer radiometric response function

The hyperspectral imager provides a high degree of flexibility in adapting the sensor to a specific measurement problem (both in design and during operation). The diffractive spectral resolution of the instrument is 2.8 nm and is over-sampled by a factor of 4 at the array detector. The full hypercube may be reduced by averaging (in real time) over adjacent spectral rows, down to the spectral resolution required for the experiment. This permits customized attention to specific spectra, reduces the data to a more manageable volume, and enables faster sampling. The averaging process also improves the signal-to-noise ratio of the instrument. The repetitive nature of the scanning process, along with remote control over the illumination and data mining operations made this instrument a good match to the robotic experimental design.

**Calibration**
Pre-scan calibrations provide spectral calibration data to convert CCD columns to wavelength, as well as dark frame and normalization frame data that are useful in converting measured radiance to reflectance. Sample spectral calibration data are presented in Figure 4. These curves are computed using the known emission lines of gas discharge lamps - in this case Xenon, Krypton, Mercury, and Neon lines. The two curves on Figure 4 are derived from orienting the array detector to measure dispersion along either the short (square symbols) or long (triangle symbol) dimension of the rectangular array detector. The long dimension is the one used in this application, providing spectral coverage from about 400 to over 1000 nm (over 780 columns binned to 130 columns). When spectral compression is used, the slope of the wavelength calibration curve of Figure 4 is multiplied by the compression factor, and the y-intercept is slightly shifted as a result.

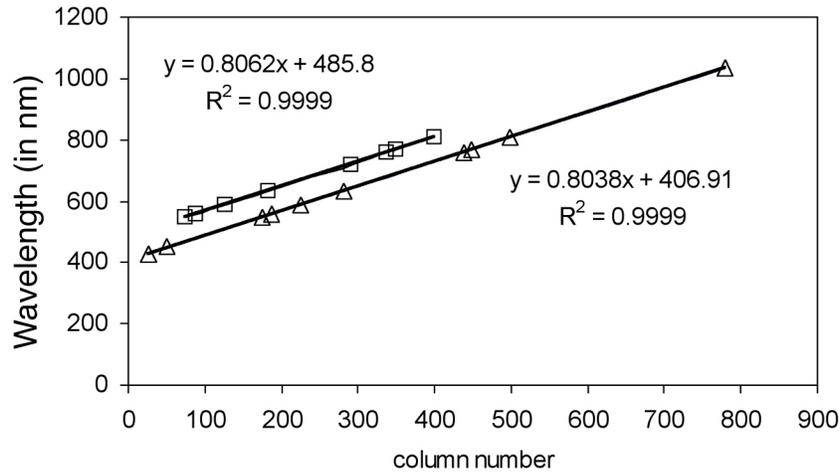

Figure 4. Spectral calibration curves for measurements of the Krypton emission lines for orthogonal orientations of the array detector. Squares are for the short dimension and triangle symbols for the long dimension of the rectangular detector (the long dimension was used in the field instrument).

We used a Krypton lamp as a test target to check the spectral linearity and accuracy of our instrument. We verified that the spectrometer bands were accurate to within 5nm across the full range of the instrument, which was deemed sufficient for the purposes of the investigation. A sample normalization frame acquired over a fully illuminated, Spectralon reflectance target is presented in Figure 5. Note the high correlation with the calculated response curve of Figure 3b. Some vignetting effects may be seen in the upper and lower parts of the frame but are limited to intensity error only, since the emission lines are straight to within one pixel variation across the detector. This is a systematic

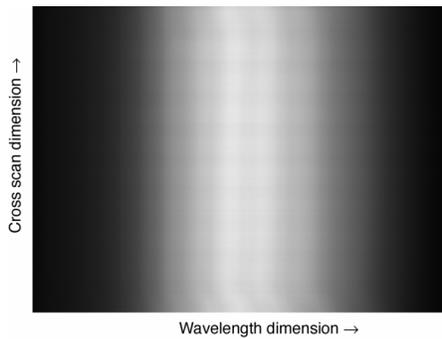

Figure 5. Measured sample normalization frame of a fully illuminated Spectralon plate.

error which may be corrected with the previously described radiometric calibration procedure.

The reflectance value of the calibration target ($R_{cal}$) may be adjusted to maximize the dynamic range of the spectrograph over the reflectivity range of the samples. For this experiment, an $R_{cal}$ of 75% was used. With this approach, when portions of the sample reflect greater than the $R_{cal}$ value the sensor will saturate and data from those pixels will be compromised. A current-controlled (less than 0.5% variation) tungsten-halogen light source was chosen to provide high reflected light levels (particularly in the IR) and stable illumination. Camera settings are optimized to provide the desired gain and integration time (generally as long as possible without causing excessive scan time) to yield maximum intensities reflected from the calibration target just under the saturation level of the camera.

Converting the raw digital number values from the camera to reflectance involves use of equation 1 over every element of the 3-D hypercube array.

$$R_{Measured} = R_{cal} \cdot \frac{(DN_{Measured} - DN_{DarkCal})}{(DN_{WhiteCal} - DN_{DarkCal})} \tag{1}$$

where DN stands for digital number (ranging 0-255 for a camera in 8-bit mode) and $DN_{WhiteCal}$ and $DN_{DarkCal}$ are the values in the corresponding pixel locations in the reference and dark calibration frames. This simple data processing approach relies on stability of the sensitivity of the CCD (typically requiring stable temperature), and uniform lighting conditions over the period from calibration to measurement.

## ANALYSIS OF HYPERSPECTRAL IMAGES

**Crystal Field Theory Background**

Crystal Field Theory is a method by which the electron energy gaps can be estimated from knowledge of the structure of a mineral, and was pioneered by Hans Bethe [*Bethe*, 1927]. Crystal field theory gives us a qualitative understanding of infrared spectra of transition metal bearing compounds in the 400-1000nm wavelength range. Transition metals are those elements with a partly filled *d* or *f* shell. Several absorption bands due to *d* shell electron promotion in transition metals are present in the 400-1000nm wavelength range [*Hunt*, 1977].

For this study, we are principally concerned with ferric ($Fe^{3+}$) iron bearing minerals – specifically hematite ($\alpha$-$Fe_2O_3$) and goethite ($\alpha$-FeOOH). An isolated, unexcited $Fe^{3+}$ ion has 23 electrons and a $3d^5$ electronic configuration – it requires five electrons to complete its *d*-shell. A *d*-shell has 5 orbitals – ($d_{xy}$, $d_{yz}$, $d_{xz}$, $d_{x2-y2}$ and $d_{z2}$) and in a free $Fe^{3+}$ ion, a *d*-shell electron has an equal probability of occupying each *d* orbital (the orbitals are degenerate). The five orbitals are broken into two symmetry groups – $t_{2g}$ for $d_{xy}$, $d_{yz}$ and $d_{xz}$ and $e_g$ for $d_{x2-y2}$ and $d_{z2}$. When a transition ion is in a crystal structure, the effect of a non-spherical electrostatic field lowers this degeneracy and splits the *d*-orbitals into different energies. The exact splitting depends on type, positions and symmetry of ligands surrounding the transition metal [*Burns*, 1993].

When a transition ion is octahedrally coordinated, the two $e_g$ orbitals point towards neighbouring ligands and therefore are less stable than the three $t_{2g}$ orbitals (this situation is reversed in cubic structures, for example).

The ground state $Fe^{3+}$ ion in an octahedral crystal field can be represented as $^6A_{1g}$. The superscript 6 designates the spin-multiplicity of this state (a sextet in this case), which is obtained by adding one to the number of unpaired electrons. Thus $^6A_{1g}$ is a sextet state with five unpaired electrons with spins aligned parallel to each other (Figure 6). Spin-allowed transitions occur between two states with the same spin multiplicity. All excited states of $Fe^{3+}$ have lower spin-multiplicities than the ground state, leading to weak 'spin-forbidden' transitions. Thus an octahedral $Fe^{3+}$ transition from $^6A_{1g} \rightarrow {}^4T_{2g}$ is a 'spin-forbidden' transition, and a lot weaker than the octahedral $Fe^{2+}$ spin allowed transition $^5T_{2g} \rightarrow {}^5E_g$, for example.

$Fe^{3+}$ in goethite, nontronite, lepidicrocite and jarosite is octahedrally coordinated. Consequently, each of these minerals has broad bands centered near 640 and 900nm corresponding to the energies for $^6A_1 \rightarrow {}^4T_2$ and $^6A_1 \rightarrow {}^4T_1$ spin-forbidden crystal field transitions (Figure 6). $Fe^{3+}$ in hematite is also octahedrally coordinated, however, it displays a greater intensity at two additional bands at 430 and 550nm due to trigonally distorted octahedral sites. This causes intensification of the $^6A_1 \rightarrow {}^4A_1$ spin-forbidden band at 430nm, and also intensifies a electron pair transition $2[^6A_1 \rightarrow {}^4T_1]$ at 550nm [*Burns*, 1993]. It is the combination of the 430 and 550nm absorption bands in hematite that give a 'flat line' in this region, rather than the high shoulder observed in goethite spectra (Figure 6).

Scheinost et al. [1998] pointed out that because of overlapping band transitions, goethite, maghemite and schwertmannite could not be discriminated by VNIR spectroscopy. It is only because we have XRD measurements showing goethite in the MARTE cores that we claim to have discriminated goethite in particular.

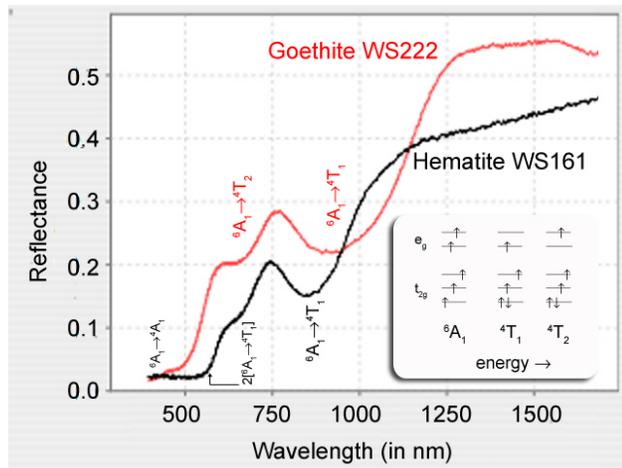

Figure 6. VNIR Spectra of Hematite and Goethite from the USGS Mineral Library [*Clark, et al.*, 2003] and crystal field theory energy diagrams for the three lowest energy spin-orbital states of tetrahedrally coordinated $Fe^{3+}$. Octahedrally coordinated $Fe^{3+}$ spin orbital states are the same, but require the addition of a g subscript to designate centrosymmetric coordination sites.

**Data Analysis Approach**

During the MARTE two week remote science operations period, the imaging spectrometer was only used to extract average spectra from small areas of core. In order to simulate limited data bandwidth conditions, the remote team could only select small regions (up to 10x10 pixels) where they could retrieve imaging spectrometer data. The remote team took some time to adapt to the user interface and only the simplest of data analysis was carried out. Simple band ratios of spectra allowed limited mineralogical interpretation. Small images were reconstructed towards the end of the two-week period that showed some promise for Fe-oxide identification.

After the two-week science operations period, the complete dataset for all the drilled cores was made available to the remote team to analyze using any means available to them. This allowed more complex methods of data analysis to be applied to the data. In particular, the data analysis took place within a prototype spectral analysis suite called "MR PRISM" [*Brown and Storrie-Lombardi*, 2006]. MR PRISM was originally designed to analyze CRISM data [*Murchie, et al.*, 2004] but is flexible enough to analyze any hyperspectral imaging datasets.

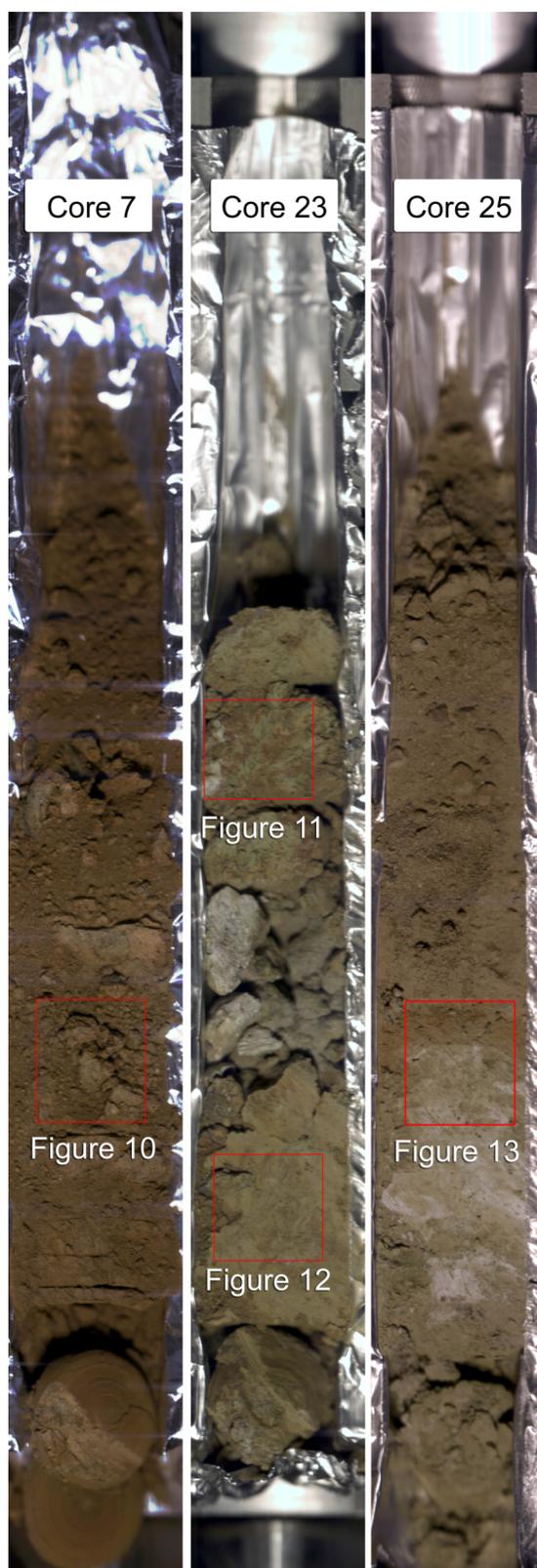

Figure 7. Approximate true color images (R=660nm, G=616nm, B=567nm) of cores 7, 23 and 25, showing locations mapped in Figures 9-12. Each core is 27mm in diameter (AQ series diamond drill core).

MARTE image cubes were delivered as 5221 separate 'frame' files for one core. Each frame was 580 pixels wide and covered 130 bands, or channels, from 413-1035nm. To analyze the cubes, a full cube was reconstructed in MR PRISM for each core. With each data point represented by 4 bytes floating point numbers, each reconstructed cube was over (5221x580x130x4bytes) 1.57Gb in size – a formidable task for data analysis.

MR PRISM was developed to analyze hyperspectral data cubes in a variety of ways, but the chosen approach for this project was to use an algorithm for absorption band mapping developed for mapping hydroxyl minerals in the Short Wave Infrared [*Brown, et al.*, 2004; *Brown, et al.*, 2005; *Brown, et al.*, 2006]. The modeling approach is discussed further below. This algorithm had not been applied to the VNIR part of the spectrum previously.

In order to keep the data analysis task manageable, three of the most morphologically interesting drill cores were selected and subsections of these drill cores were chosen where interesting mineral textures and possible compositional gradients were thought to exist. The cores chosen were 7, 23 and 25. Approximate true color images of these cores appear in Figure 7. Our interpretation of the spectra relied heavily on crystal field theory.

**Data Analysis Technique**

We follow an absorption band curve fitting method which has previously been applied successfully in the 2000-2500nm range [Brown, 2006]. The steps of the process are outlined in Figure 8. First, the straight line continuum is removed (subtracted) from the spectrum [Clark, et al., 1987], then converted to energy (wavenumber) space [Sunshine and Pieters, 1993], then Gaussian curves are generated to match the resultant peaks. The Gaussian parameters are improved in an iterative least squares fitting procedure, until a sufficiently precise match is obtained. We measured errors using a least squares chi-squared value. Each iteration the chi-squared value is compared to a target value that was computed based on the number of free parameters in our fit. If the difference between chi-square and target value was below a tolerance value (0.001), we considered the least squares run to have converged. In this way, our analysis resembles that described by Press [1989]. The resultant Gaussian curves are converted back into wavelength space, and then analyzed using a rule based expert system in order to produce a mineral map. Parameters of the Gaussian curves such as amplitude, centroid and width are stored and may be mapped over the processed image to obtain greater insight into the spectral properties of each pixel.

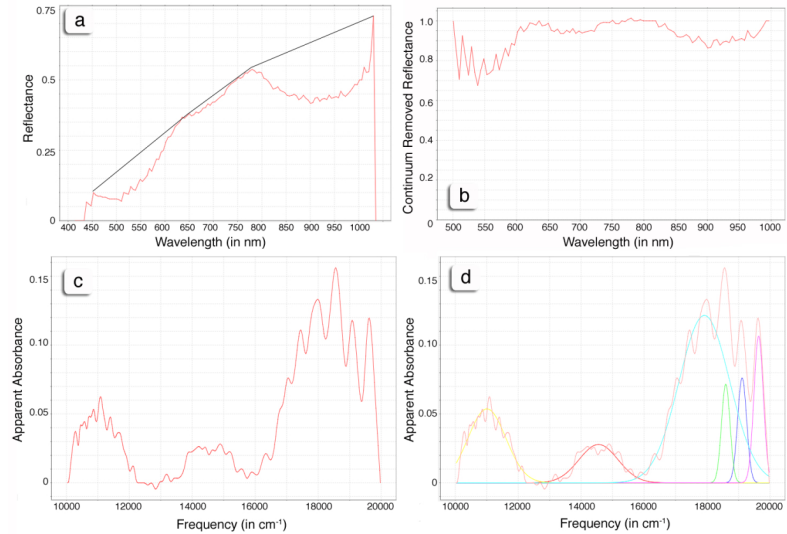

Figure 8. Summary of the curve fitting process. (a) Original spectrum and straight line continuum, (b) continuum removed spectrum (subset of original spectrum from 500-1000nm), (c) inverted wavenumber (cm$^{-1}$) space curve, (d) fitted wavenumber space curve. Note the colored Gaussian curves have captured all absorption bands present.

The data analysis technique employed here attempts to model the $Fe^{3+}$ absorption bands in hematite and goethite, since these are the primary minerals occurring in the MARTE drill core. Crystal field theory and experimental data suggest that hematite will have a shorter wavelength for the $^6A_1 \rightarrow {}^4T_1$ transition than goethite [*Bishop, et al.*, 1993], and the $^6A_1 \rightarrow {}^4T_2$ transition absorption band will be obliterated in hematite due to an intense paired transition $2[^6A_1 \rightarrow {}^4T_1]$ (Figure 6). Thus the rule-based system we developed classified a mineral as 'Hematite' if it possessed a $^6A_1 \rightarrow {}^4T_1$ transition less than 900nm, and goethite if it possessed a $^6A_1\rightarrow{}^4T_1$ transition greater than 900nm, in combination with a $^6A_1\rightarrow{}^4T_2$ transition at around 690nm.

In order to learn more about the positions of absorption bands in every analyzed image, our software program also produced a 'curve fitting histogram' that reported the central points of every recognized absorption band in the image. These are summed to create a histogram, and have been reported below. The advantage of this approach is that if an unrecognized absorption band is appearing in many of the pixels in the image, it will become apparent as an anomalous peak. We found during the 'fine tuning' of our curve fitting parameters, it was invaluable to have quantitative information on where most absorption band peaks were being found. This allowed us to make decision on whether to adjust our parameters (and thus include more identified pixels in a mineral map).

It is important to note that previously in the literature, authors have reported the central wavelength of $^6A_1\rightarrow{}^4T_1$ and $^6A_1\rightarrow{}^4T_2$ features as the minimum of non-continuum-removed spectra. Due to the steep slope of most spectra in the 500-700nm region, when the continuum is removed, the absorption band center shifts to the right (to higher wavelengths). Thus the figures reported in this paper are approximately 30-40nm higher than those reported previously in the literature. We have decided to take this departure from custom based on the following line of reasoning:

1. All VNIR spectra, particularly those in the visible, are affected by a strong drop-off in reflectance due to absorption by transition metals.
2. It is difficult to know the true strength of this absorption and model its effects as a Gaussian shape, because the central absorption point of such a shape does not exist. The relevance of Gaussian modeling of this continuum dropoff is therefore unjustified at best unclear, and this has not been attempted to our knowledge.

3. A straight-line (or linear) continuum model of the effects of the strong transition ion dropoff is the most relevant way to compensate for an absorption of variable strength.

4. It is essential to make some compensation for this strong continuum absorption because of its variability, in order to allow the comparison of weaker bands (such as those discussed in this paper).

We therefore believe that studies that have simply stated the minimum of the band (by eye) are not easy to compare with other results that have differing and uncompensated continuum slopes.

## XRD VALIDATION

X-ray diffraction analysis was performed on powdered samples (< 45 μm) extracted from cores 23 and 25 after the mission simulation. Core 7 was not analyzed with XRD. Samples were examined by a PANalytical X'Pert Pro PW 3040/60 diffractometer run under CoKα, at 40kV, 40mA, with a scan range 4-80 2θ, step of 0.02 and at 20s per scan. Goethite and hematite were the dominant Fe-bearing minerals detected in cores 23 and 25. White material found in both cores was attribute to mica – either as illite or muscovite. Trace amounts of kaolinite and smectite were also detected in core 23 but not core 25 [*Sutter, et al.*, 2007].

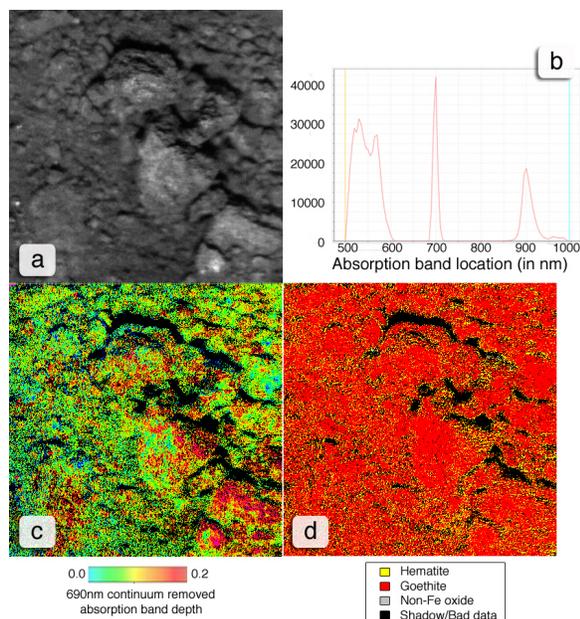

Figure 9. Core 7 subset analysis results. All images approximately 18mm across. (a) Greyscale image of channel at 567nm. (b) Curve fitting histogram showing the cumulative total of all absorption bands found. (c) 690nm absorption band depth map (d) Mineral classification map.

## RESULTS

**Core 7**

Core 7 consisted mostly of fine, ochre-red soil, with little visible compositional variety (Figure 7). Some clumps of material and small pebbles were present. One 400x400 pixel subset of Core 7 was chosen for analysis in proximity to a number of small pebbles in the bottom part of the core. The results are shown in Figure 9. From the mineral map, it can be seen that the scene is full of goethite, with small amounts of more hematitic material appearing to be sprinkled throughout the scene, with no strong preference for the consolidated material in the scene. From the 690nm absorption band depth map, it can be seen that the strongest 690nm absorption bands are associated strongly with the competent pebbles (eg. in the image lower right). It is likely that more highly crystalline goethite within the pebbles is responsible for this effect.

**Core 23**

Core 23 had the most compositional variety of any cores analyzed during the MARTE experiment. Using XRD, traces of muscovite/illite were found in 1cm wide white pebbles in the core, as well as yellow internal veining of red-purple chips [*Sutter, et al.*, 2007].

Two 400x300 pixel subsets of the core were sampled (Figure 7). The first subset captured a radially-streaked part of the core that displayed intriguing texture and color variations. The results of the analysis are shown in Figure 10. The mineral map shows that small amounts of hematite are clumped together in moderately dark albedo parts of the scene (especially the top left). This is the only core we analyzed where the iron oxide absorption bands showed strong indications of cohesive zones of hematite. The mineral mapping also turned up several points on the left hand side of the image showing no absorption bands, and hence no iron bearing minerals. These locations show high albedo and were sampled and analyzed by XRD – most of them contain muscovite/illite [*Sutter, et al.*, 2007]. The 690nm absorption band depth map in Figure 10 shows the strongest absorption bands appear to form linear streaks going from bottom left to top right. This is in accord with the visible veining relationships in this part of the core.

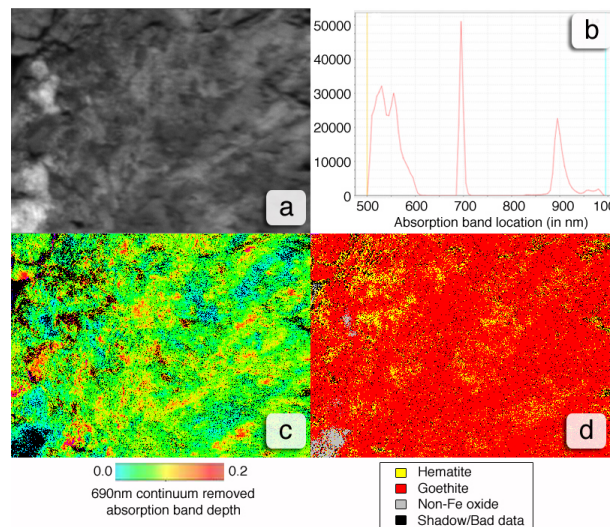

Figure 10. Core 23 subset 1 analysis results. All images approximately 18mm across and 14mm high. (a) Greyscale image of channel at 567nm. (b) Curve fitting histogram showing the cumulative total of all absorption bands found. (c) 690nm absorption band depth map (d) Mineral classification map.

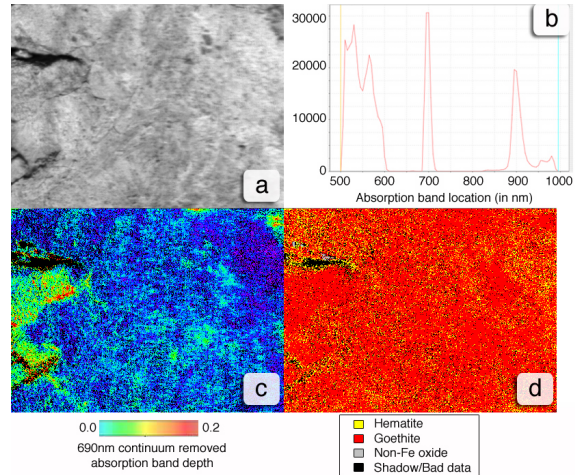

Figure 11. Core 23 subset 2 analysis results. All images approximately 18mm across and 14mm high. (a) Greyscale image of channel at 567nm. (b) Curve fitting histogram showing the cumulative total of all absorption bands found. (c) 690nm absorption band depth map (d) Mineral classification map.

The second 400x300 pixel subset of Core 23 that was analyzed appears in Figure 11. The mineral map shows that this area is again predominantly goethitic. No patches of hematite are present here, but there is an intriguing occurrence of a non Fe-oxide of low albedo in the top left quadrant of the image (above the shadow). It is not just a shadow or crack because even shadows containing Fe-oxide show evidence of Fe-oxide bands. It is probably not muscovite, due to its relatively low albedo, but we were unable to sample this point with XRD. The absorption band map shows that even though the whole surface of this competent sample is goethitic, the strongest absorption bands occur in a spatially coherent part of the surface on the left hand side of the image. It is likely the sharp contact between this high absorption band area and the medium absorption band region is due to a high crystalline matrix in this inclusion.

**Core 25**

Core 25 displayed an intermediate rock-soil nature to cores 7 and 23. It contained some consolidated material, chiefly in the bottom part of the core (Figure 7).

We chose to analyze a section of core 25 that displayed a transition from disaggregated material to more consolidated material. The section analyzed is shown in Figure 12. The mineral map shows almost all the scene is goethite, but also that there are regions of the more consolidated material that are not Fe-oxide. These appear in the bottom part of Figure 12. The regions appearing to have no Fe-oxide bands are also high in albedo. Similar regions were sampled in core 25, and displayed a 50/50 mixture of goethite and muscovite/illite [*Sutter, et al.*, 2007]. The areas of the consolidated material that were not high in albedo display relatively deep 690nm absorption bands, suggesting more crystalline goethite may be present in lower albedo parts of the consolidated material.

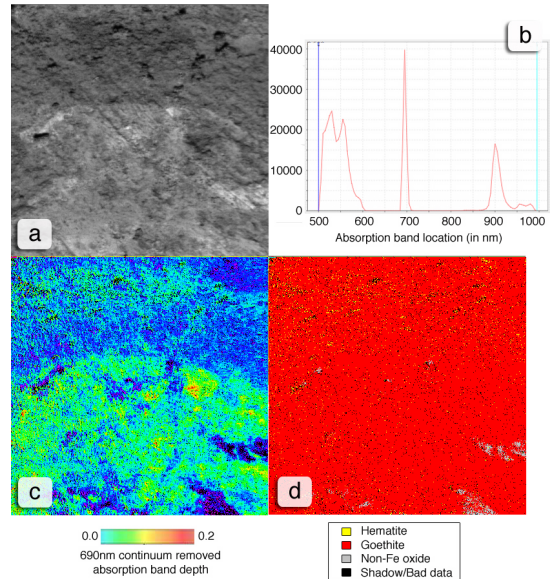

Figure 12. Core 25 subset analysis results. All images approximately 18mm across. (a) Greyscale image of channel at 567nm. (b) Curve fitting histogram showing the cumulative total of all absorption bands found. (c) 690nm absorption band depth map (d) Mineral classification map.

## DISCUSSION

**Interpretations**

The observations of the mineral and absorption band maps shown in Figure 9-12 suggest that most of cores 7, 23 and 25 were dominantly goethitic, with small pockets of hematite and non Fe-oxide material spread throughout parts of the competent core. It is interesting that the most spatially coherent hematite cells occurred in association with a high albedo non Fe-oxide we found to be muscovite/illite [*Sutter, et al.*, 2007]. One possibility is the hematite may have formed from goethite as an oxidative by product in a less-reducing hydrothermal reaction or hematite may have formed by reorganization of ferrihydrite. Note that due to the band matching technique we have adopted, in some regions where there are shadows or bad data, it is statistically more likely that hematite will be 'detected' in the edges of these areas since hematite detections only require the identification of one absorption band, whereas goethite detections require two. Thus hematite detections on the edges of 'no data' or 'shadowed' regions require careful examination.

The final core analyzed, core 25, appears to contain consolidated material that is a mixture of goethite and high albedo non-Fe oxide. From post-mission XRD analysis, we conclude the non-Fe oxide is muscovite or some other phyllosilicate material, and this material is a vestige of highly penetrative hydrothermal alteration, with goethite and phyllosilicate-rich zonal alteration regions.

This study made several new findings in the field of VNIR spectroscopy. These were mostly facilitated by the absorption band data analysis method, which is designed to explore the data and uncover hidden details when a numerical description of the spectral bands is available.

1. **Constraints on the spread of Fe-oxide absorption bands.** All three cores analyzed showed similar absorption band histograms (shown in Figures 9-13), despite the variation in soil-rock ratios and mineral composition. It cannot be taken as proven that this will always be the case, but this finding demonstrates the general 'stability' of Fe-oxide bands.

2. **Nature of the 690nm $^6A_1 \rightarrow {^4T_2}$ feature distribution.** The tight spread of the 690nm feature makes it useful for goethite detection. Scheinost et al. [1998] point out that ferrihydrite and feroxyhite are identifiable in the VNIR because both display this feature at wavelengths greater than 700nm, however we found no evidence for these minerals in our samples. We found absorption band depth maps of this feature to be extremely useful in determining regions of more crystalline goethite.

3. **Nature of the 900nm $^6A_1 \rightarrow {^4T_1}$ distribution.** We were able to use the T1 900nm feature as a compositional index, with goethite-rich regions displaying a central wavelength > 900nm, and goethite-poor regions showing a T1 feature < 900nm. This has been reported by other authors [*Cudahy and Ramanaidou*, 1997], however we have here applied it to a real life scenario with an imaging spectrometer.

4. **Ability to detect non Fe-oxide materials.** It is a challenging matter for simple (or complex) band ratios, false color images and (especially) principle components analysis to identify the presence of non Fe-oxide material in the VNIR. We have demonstrated the ability of this method to automatically discriminate both high and low albedo non Fe-bearing material using the absence of strong absorption bands.

**Experimental Limitations**

VNIR spectroscopy is inherently limited to observing the nature of transition metal electronic absorption bands. As such, the presence of silicates, phyllosilicates, amphiboles, pyroxenes, sulfates, carbonates can only largely be estimated. We consider this apparatus and its performance on the MARTE project a successful proof of concept. A future Mars drilling rig should be equipped with an instrument that can cover at least to the Short Wave Infrared (up to 2500nm) and possibly beyond.

**Future Work**

At this time we are unable to capitalize on the (possibly bimodal) distribution of absorption bands around 550nm. With our current absorption band implementation, most pixels showed at least two absorption bands between 500 and 600nm. We were unable to recognize a significant link to composition at this time, due to two factors: 1.) Fe-oxides have several overlapping bands in this region [*Scheinost, et al.*, 1998] and 2.) The MARTE Imaging Spectrometer has lower signal to noise in this part of the spectrum. Should an imaging spectrometer be selected for drill core analysis on a future Mars mission, higher signal to noise characteristics should be sought in order to allow automated band analyses such as attempted here.

## CONCLUSION

We have summarized the design and construction of the VNIR Imaging Spectrometer, and described a unique absorption band modeling method that was used for data analysis. In so doing, we have highlighted the usefulness of this technique for differentiating between goethite – hematite mixtures and for recognition of non Fe-bearing minerals. Although this information is only of indirect use in the search for life in subsurface drill cores, it can give invaluable contextual information on the mineralogical environment. A hyperspectral spectrometer covering at least the VNIR will be an essential part of a future Mars remote drilling Astrobiology mission.

## ACKNOWLEDGEMENTS

We would like to thank the entire MARTE Team, especially the PI Dr. Carol Stoker, for their commitment and camaraderie during the project. We would also like to thank Dr. Ted Roush and two anonymous referees for their thoughtful reviews.